\documentclass[twocolumn]{IEEEtran}
\usepackage{cite}
\usepackage{textcomp}
\usepackage{graphicx}
\usepackage[cmex10]{amsmath}
\usepackage{algorithmic}
\usepackage{array}
\usepackage{url}
\newcolumntype{C}[1]{>{\centering\let\newline\\\arraybackslash\hspace{0pt}}m{#1}}

\begin{document}
\title{A Unified Multicarrier Modulation Framework}

\author{Konstantinos~Maliatsos\thanks{K.~Maliatsos and A.~Kanatas are with the Department
of Digital Systems, University of Piraeus, Piraeus, Greece (e-mail: \{kmaliat,kanatas\}@unipi.gr).}, Eleftherios Kofidis\thanks{E.~Kofidis is with the Department of Statistics and Insurance Science, University of Piraeus, Piraeus, Greece and the Computer Technology Institute \& Press 
``Diophantus'' (CTI), Greece (e-mail: kofidis@unipi.gr).}, and~Athanasios~Kanatas
}

\maketitle

\begin{abstract} 
Orthogonal frequency division multiplexing (OFDM) has been recently recognized as inadequate to meet the increased requirements of the next generation of communication systems. A number of alternative modulation solutions, based on the use of filter banks, have thus been proposed and are currently being considered as candidate waveforms for the envisaged air interface of the future networks. A unified view of these schemes would largely facilitate a systematic comparison of their pros and cons as well as the study of methods for related signal processing problems. To this end, a generic modulator is developed in this paper, following a structured matrix formulation. With appropriate parameter settings, existing modulation schemes can result as special cases. Three such popular examples are presented in detail. 
\end{abstract}

\section{Introduction}

\IEEEPARstart{O}{rthogonal} Frequency Division Multiplexing (OFDM)~\cite{Li:2009:OFD:1823628} and OFDM-based transmission schemes have dominated modern wireless communications standards, including WLANs, WiMAX, LTE, DVB. Despite its simplicity and its proven effectiveness, the OFDM modulation also presents significant weaknesses, which include reduced spectral (and power) efficiency due to the cyclic prefix redundancy and low tolerance to frequency offset/synchronization errors and narrow band interference, which can be limiting factors in un-coordinated multi-user environments such as cognitive radios~\cite{zlrms10}. It is for these (and other) limitations that other candidate modulation formats are considered for the future mobile networks~\cite{heir}.

Internet of Things (IoT) applications and Device-to-Device (D2D) communications are expected to significantly increase the internet traffic~\cite{m15}. The transition to Machine-Type Communications (MTC) also marks a transition from the cell-centric network model to distributed modes, where spectral efficiency, high tolerance to interference and robustness to frequency-time synchronization errors are necessary attributes for the modulation waveform. In view of these requirements, filterbank-based multicarrier (FBMC) modulation schemes have recently come (back) to the attention of the research community as a potential solution to the limitations of OFDM. Starting from the rather straightforward Filtered Multi-Tone (FMT) scheme~\cite{1007382}, many candidate waveforms have been proposed for the mobile systems of the future~\cite{f11}. Offset Quadrature Amplitude Modulation-based FBMC (FBMC-OQAM or OFDM-OQAM)~\cite{995073,Farhang} has been in the center of relevant work in major research projects (e.g., \cite{phydyas,metis,EMPhAtiC}). FBMC-OQAM retains orthogonality and optimal spectral efficiency under ideal conditions, while being more robust under interference and frequency offset errors compared to OFDM, at the cost of increased computational complexity. In addition to FBMC-OQAM, a plethora of other filterbank-based waveforms have been proposed with popular examples including the Universal Filtered Multi-Carrier (UFMC)~\cite{6824990,6900754} and the Generalized Frequency Division Multiplexing (GFDM)~\cite{5073571} schemes. For each waveform, there is a tradeoff of advantages and drawbacks per scenario that has to be evaluated. Thus, the comparative evaluation of these waveforms within a common framework would play a crucial role in defining the 5G waveform~\cite{6877912,5483521}. Moreover, there is still room for new versions of waveforms, resulting from appropriate tweaks (e.g., \cite{Lin2014,Tonello2014}), which may be advantageous in certain conditions.
  
This paper reports recent results of our work towards the development of a unified framework for multicarrier/filterbank-based modulation schemes. A general transmitter configuration is presented, which can serve as a generic modulator, able to accommodate all the best known candidate waveforms, with the appropriate parameter settings. Related existing works include~\cite{5425501,rh11,6362266,zlcp13,gls14}. A distinguishing feature of the present work is that it covers the multicarrier waveforms \emph{and} their single-carrier extensions~\cite{5577747}, in a detailed matrix formulation. The latter can be used to unify existing and develop new methods of transmit-receive signal processing in a general FBMC context. An implementation of the generic modulator has been developed in MATLAB in order to verify the validity of the model and use it as a tool for evaluation of different waveforms.  

\subsection{Notation}

Vectors and matrices are denoted by bold lower- and upper-case letters, respectively. The superscript $^{*}$ stands for complex conjugation. The imaginary unit is written as $j$. For a matrix $\mathbf{A}$, $[\mathbf{A}]_{m,n}$ is its $(m,n)$ entry and $\mathbf{A}^{T}$ is its transpose. $\otimes$ and $\circ$ stand for the Kronecker and Hadamard (entry-wise) product, respectively. The identity matrix of order $N$ is denoted by $\mathbf{I}_N$. $\mathbf{0}_{r\times s}$ is the $r\times s$ matrix of all zeros. The $N\times 1$ vector of all ones is denoted by $\mathbf{1}_N$. 

\section{General Description}

\begin{figure}
\centering
\includegraphics[width=9cm]{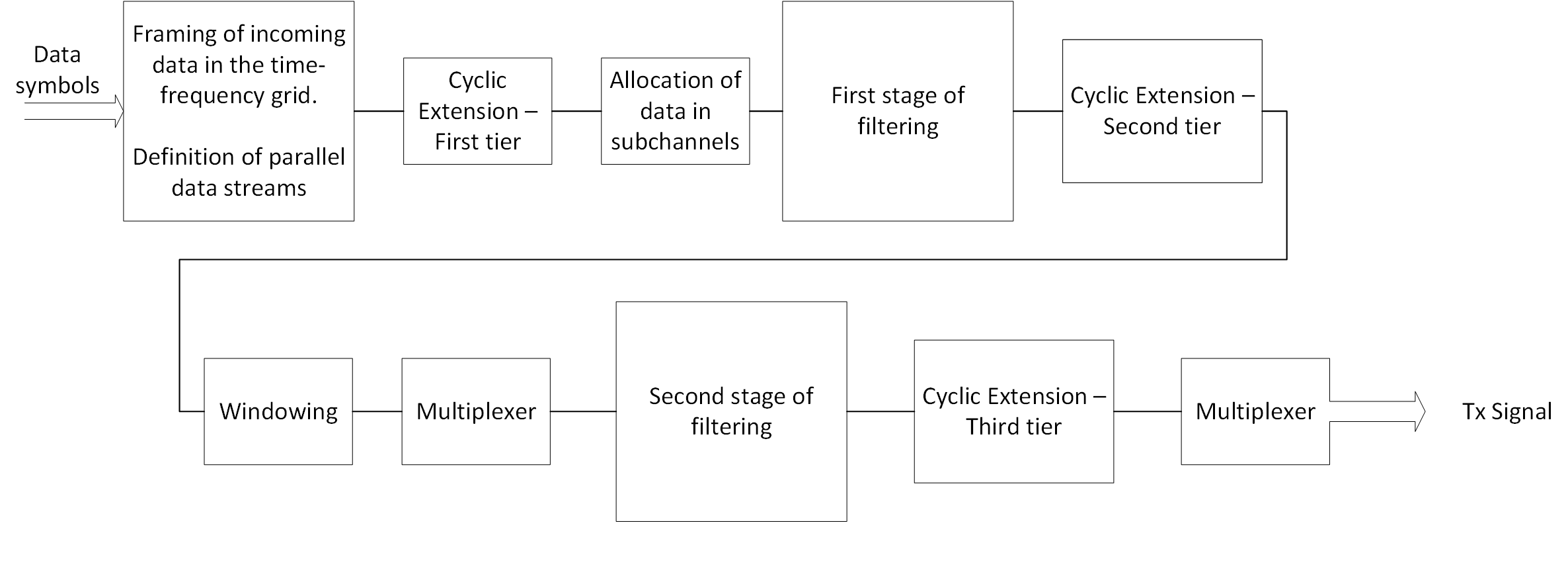}
\caption{A block diagram of the generic modulator.}
\label{Fig1}
\end{figure}

A block diagram of the generic modulator is presented in Fig.~\ref{Fig1}. 
It must be emphasized that this work is not concerned with issues of computationally efficient implementation. It instead focuses on developing a framework that is able -- with proper parametrization -- to result in any of the known multicarrier/filterbank-based modulation schemes. Of course, not all the blocks are included in any particular modulation type. With proper selection of matrices and setting of parameters, some blocks may be canceled / omitted when yielding the final representation of a given modulation scheme.

The basic blocks of the generic modulator are the following:
\begin{itemize}
	\item \emph{Data Framing}: In this block, the data symbols are organized as data frames in the time-frequency grid. Moreover, and depending on the transmission scheme, it is possible to synthesize parallel time-frequency data grids that are transmitted simultaneously. The data framing procedure is necessary for all modulation schemes.
	\item \emph{Cyclic Extension}: There are three cyclic extension blocks. Multiple tiers for cyclic extension of the signal are required in order to cover all possible cases. For example, FBMC modulation with cyclic prefix requires signal extension directly in the data frame, while in OFDM, cyclic prefix is added after the inverse DFT (i.e., after a filtering operation). Finally, in SC-FDMA and UFMC, a cyclic prefix is added at the end of the transmission chain. It is possible to cancel one or more cyclic extension blocks by setting the corresponding prefix/suffix length(s) equal to zero. 
	\item \emph{Generalized Filtering}: The term ``Generalized Filtering" is used because, besides the convolution of the signal with a filter, this block also performs rate conversion, phase correction and subchannel modulation. Two filtering stages are needed in order to model waveforms with multiple subchannelization tasks, e.g., UFMC and SC-OFDM/SC-FBMC. Moreover, circular convolution instead of linear is assumed, so that waveforms like those in~\cite{Lin2014,Tonello2014} can be integrated. Linear convolution can be achieved by appropriately setting the circular filtering period. To omit a filtering stage, the corresponding impulse response is assumed to be ideal while no rate conversion is performed.     
	\item \emph{Windowing}: The windowing block is used for time-domain signal windowing (e.g., in Windowed OFDM~\cite{5764431}), which is used in most cases to suppress sidelobes. If no windowing is necessary, then a vector of all ones (ideal) is considered.
	\item \emph{Commutator - Multiplexing}: These blocks interconnect the filtering stages and also generate the final waveform as output. They are used to re-allocate and combine subchannels before and after the filters. 
\end{itemize}
The following sections are devoted to describing these blocks in detail. 

\section{Signal Input}
\label{Sec1}

At first, the incoming data are organized in data frames. The dimensions of a data frame are assumed to be
$N \times {M'}$, where $M'$ is the number of frequency subchannels currently occupied by the modulator and   
$N$ is the number of symbols per subchannel and per frame in the time domain. Any input data constellation (e.g., QAM, PSK, PAM, etc.) may be employed. 

In many cases, multiple data frames can be simultaneously transmitted. For example, in FBMC-OQAM, two parallel streams of real-valued symbols (PAM) are considered. Moreover, with MIMO transmission, multiple streams of data may be combined to produce signals for the transmit antennas. Assuming that $P$ signals can be sent in parallel, the input signal frame can be expressed as a function of three variables ${x_{p,m,n}}$, $p = 0,1,\ldots,P - 1$, $m = 0,1,\ldots,M' - 1$, $n = 0,1,\ldots,N - 1$ or equivalently as a $PN \times {M'}$ matrix ${\bf{X}} = {\left[ {\begin{array}{*{20}{c}}
{{\bf{X}}_0^T}&{{\bf{X}}_1^T}& \cdots &{{\bf{X}}_{P - 1}^T}
\end{array}} \right]^T}$ 
with $\mathbf{X}_p$ being the $N \times M'$ matrix of the $p$-th parallel channel.

If the number of available data,  ${N_\text{data}}$, exceeds $M'N P$, then more than one input signal frames are defined and thus a superscript indicating the input signal frame index should be added, as in
\begin{equation}\label{eq1}
x_{p,m,n}^{(i)},\; i = 0,1,\ldots,\left\lceil {\frac{{{N_\text{data}}}}{{M'NP}}} \right\rceil  - 1
\end{equation}
It is noted that in the general case where $\frac{N_{\text{data}}}{M'NP}$ is not an integer, the input signal should be properly zero-padded and probably scrambled, depending on the waveform properties.

\subsection{First cyclic extension: directly on the data frame}
\label{SubSec11}

The first possible choice for the application of signal cyclic extension is directly on the input data frame. Let the number of time samples of the input signal frame after the insertion of the prefix be denoted by $N^{(1)}_{s}$. In this general framework, it is also possible to add a zero prefix or suffix. If ${N^{(1)}_\text{zp}}$, ${N^{(1)}_\text{cp}}$, ${N^{(1)}_\text{cs}}$, and ${N^{(1)}_\text{zs}}$ are the lengths of the zero prefix, the cyclic prefix, the cyclic suffix, and the zero suffix, respectively, then $N^{(1)}_s = {N^{(1)}_\text{zp}} + {N^{(1)}_\text{cp}} + N + {N^{(1)}_\text{cs}} + {N^{(1)}_\text{zs}}$, which is the wideband input signal period. If $N^{(1)}_s = N^{(1)}_\text{cp} + N^{(1)}_\text{cs} + N$, then no zero prefix/suffix is introduced between consecutive data frames. If $N^{(1)}_s>N^{(1)}_\text{cp} + N^{(1)}_\text{cs} + N$, then zero guard is inserted between frames, while if $N^{(1)}_s< N^{(1)}_\text{cp} + N^{(1)}_\text{cs} + N$, then there is overlap of the data frames in the time domain.

Generally, the $i$-th input signal frame with the addition of the cyclic extensions can be expressed as in~(\ref{xpmni}) (see top of the next page)
\begin{figure*}
\begin{equation}
{x'}_{p,m,n}^{(i)} = \left\{ \begin{array}{ll}
  0, & n =  - N^{(1)}_\text{zp} - N^{(1)}_\text{cp},\ldots,- N^{(1)}_\text{cp} - 1 \\ 
  x_{p,m,\bmod (n,N)}^{(i)}, & n =  - N^{(1)}_\text{cp},\ldots,N + N^{(1)}_\text{cs}-1 \\ 
  0, & n = N + N^{(1)}_\text{cs},\ldots,N + N^{(1)}_\text{cs} + N^{(1)}_\text{zs} - 1
\end{array} \right.
\label{xpmni}
\end{equation}
\hrulefill
\end{figure*}
The overall input signal per subchannel, per parallel stream and for multiple frames can then be expressed (in the time domain) as
\begin{equation}\label{eq3}
{{x'}_{p,m,n}} = \sum\limits_{i = 0}^{\left\lceil {\frac{{{N_{\mathrm{data}}}}}{{MNP}}} \right\rceil  - 1} {{x'}_{p,m,\bmod (n - i{N^{(1)}_{s}},N)}^{(i)}} 
\end{equation}
In the following, each individual data frame will be considered separately for the extraction of the matrix relationships. Hence the data frame index $i$ will henceforth be omitted from the presented matrix expressions.

Algebraically, the prefix/suffix insertion procedure can be expressed as a matrix multiplication with the following matrix
\begin{equation}\label{eq4}
{\mathbf{C}^{(1)}} = \left[ {\begin{array}{*{20}{c}}
  {{{\mathbf{0}}_{{N^{(1)}_\text{zp}} \times N}}} \\ 
  {\begin{array}{*{20}{c}}
  {{{\mathbf{0}}_{{N^{(1)}_\text{cp}} \times \left( {N - {N^{(1)}_\text{cp}}} \right)}}}&{{{\mathbf{I}}_{{N^{(1)}_\text{cp}}}}} 
\end{array}} \\ 
  {{{\mathbf{I}}_N}} \\ 
  {\begin{array}{*{20}{c}}
  {{{\mathbf{I}}_{{N^{(1)}_\text{cs}}}}}&{{{\mathbf{0}}_{{N^{(1)}_\text{cs}} \times \left( {N - {N^{(1)}_\text{cs}}} \right)}}} 
\end{array}} \\ 
  {{{\mathbf{0}}_{{N^{(1)}_\text{zs}} \times N}}} 
\end{array}} \right]
\end{equation}
Indeed, the extended $p$th parallel data stream can then be written as
\begin{equation}\label{eq5}
{{{\mathbf{X}}'}_p} = {\mathbf{C}^{(1)} \mathbf{X}_p}
\end{equation}
For $P$ parallel streams, the following block diagonal matrix applies:
\begin{equation}\label{eq6}
{{\mathbf{C}_{\mathrm{ext}}^{(1)}}} = \text{diag}\underbrace{(
  \mathbf{C}^{(1)},\mathbf{C}^{(1)},\ldots,\mathbf{C}^{(1)})}_{P~\text{times}}=\mathbf{I}_P\otimes \mathbf{C}^{(1)}
\end{equation}

This procedure concludes the description of the input signal. Specific data symbol design and parameter selection may lead to different well-known modulations.

\section{First Generalized Filtering Stage}
\label{Sec2}

This section describes the first filtering stage of the generic modulator. The filtering stage performs: a)  filtering/shaping/channelization , b) rate conversion, c) time offset insertion, and d) phase offset correction.

Filtering is performed with $M_1$  subchannels where $M_1 \ge M'$. Subchannels not carrying an input signal do not contribute to the composition of the total transmitted signal. Thus, the first component of the first filtering stage is a commutator that corresponds input signal frame subchannels into the filters. The signal in the input of the filters is given by:
\begin{equation}\label{eq7}
{x''}_{p,{{[{\mathbf{e}}]}_m},n}^{(i)} = {x'}_{p,m,n}^{(i)}
\end{equation}
where $\mathbf{e}$ is an $M' \times 1$ commutator vector that matches the input signal frame subchannels with the filter subchannels and therefore ${\left[ {\mathbf{e}} \right]_m} \in \{0,1,\ldots, M_1 - 1\}$, $m = 0,1,\ldots,M' - 1$.

This step can be expressed in matrix form by defining the $M' \times M_1$ 0-1 commutator matrix $\mathbf{E}^{(1)}$ as follows. If the $m$th data frame subchannel is filtered by the $k$th filter, then ${\left[ {\mathbf{E}^{(1)}} \right]_{m,k}} = 1$, while ${\left[ {\mathbf{E}^{(1)}} \right]_{m,l}} = 0$ for $l\neq k$, $l = 0,1,\ldots,M_1 - 1$. Therefore, each row of $\mathbf{E}^{(1)}$ has exactly one entry equal to unity with the rest of them being zeros. 
Commutation is then performed by post-multiplying the data frame matrix $\mathbf{X'}_p$ with $\mathbf{E}^{(1)}$, i.e., $\mathbf{X''}_p= 
\mathbf{C}^{(1)}\mathbf{X}_p\mathbf{E}^{(1)}$.
Moreover, $\mathbf{E}^{(1)}$ can be directly applied to the extended model for the $P$ parallel streams, i.e., $\mathbf{X''}= 
\mathbf{C}_{\mathrm{ext}}^{(1)}\mathbf{X}\mathbf{E}^{(1)}$.

Filtering can be combined with rate conversion. Thus, the filtering procedure consists of an upsampling-convolution-downsampling chain. It is assumed that the signal is upsampled by $L_1$ and downsampled by $Q_1$ before and after the convolution procedure respectively.  

\subsection{Upsampling and convolution}
\label{SubSec21}

Since linear filtering can be expressed as circular filtering of infinite period, circular filtering is assumed. This is because in some variations of multicarrier modulations circular filtering is employed~\cite{Lin2014,Tonello2014}. It is noted that circular filtering is not directly related to signal cyclic extension, that is, it is not used to emulate the insertion of the cyclic extension. Thus, cyclic extension may or may not co-exist with circular filtering in a modulation scheme. Let the period of the circular filtering be $N_{\mathrm{c}}^{(1)}$. For linear filtering, $N_{\mathrm{c}}^{(1)}$ has a value large enough so that linear and circular filtering results coincide, i.e., $N_{\mathrm{c}}^{(1)} \ge \mathrm{data~frame~length}+K_0-L_1$, where $K_0$ is the filter length for each subchannel.\footnote{In case the subchannel filters are not modulated versions of the same prototype filter, ${K_0}$ can be taken as the maximum of their lengths.}  
Moreover, the following quantities are defined:
\begin{itemize}
	\item Filter coefficients $h_n^m$ for the $m$-th subchannel. Each filter is considered as a lowpass (baseband) FIR filter. Therefore, modulation of the filter is required to place the signal at the desired subband.
	\item Duration $N_{\mathrm{c}}^{(1)}$ (in samples) of the filtered wideband frame  the highest, up to this point, processing rate. As mentioned above, this also indicates the circular filtering period if circular filtering is assumed.
	\item The time offset, ${o^{(1)}_{p}}$, in samples, for each parallel stream $p = 0,1,\ldots,P - 1$.
\end{itemize}
The superscript ${(1)}$ indicates that the defined parameters relate to the 1st filtering stage.

Upsampling and circular filtering with proper spectral positioning of a given input signal and a given filter can be expressed as in~(\ref{eq9}) (see top of the next page).
\begin{figure*}
\begin{equation}\label{eq9}
{x'''}_{p,{{[{\mathbf{e}}]}_m},n}^{(i)} = \sum\limits_{u = \left\lceil {\frac{{\bmod (n - {o^{(1)}_p},{N_{\mathrm{c}}^{(1)}}) - {K_0} + 1}}{L_1}} \right\rceil }^{\left\lfloor {\frac{{\bmod (n - {o^{(1)}_p},{N_{\mathrm{c}}^{(1)}})}}{L_1}} \right\rfloor } {h_{\bmod(n - {o^{(1)}_p} - L_1u,{N_{\mathrm{c}}^{(1)}})}^{{{[{\mathbf{e}}]}_m}}{e^{\frac{{2\pi j{{[{\mathbf{e}}]}_m}{\bmod(n - {o^{(1)}_p} - {L_1}u,{N_{\mathrm{c}}^{(1)}})}}}{{{M_1}}}}}{x'}_{p,m,u}^{(i)}}
\end{equation}
\hrulefill
\end{figure*}
The time offset $o^{(1)}_p$ is used in order to insert the OQAM functionality in the generic modulator. For example, for single-antenna OQAM modeling, $P=2$ and $o^{(1)}_0=0$, $o^{(1)}_1=M_1/2$, since in FBMC-OQAM two parallel sets of real-valued symbols are transmitted with a time offset of $M_1/2$ between them. 

To extract the equivalent to~(\ref{eq9}) matrix expressions, the upsampling matrix ${\mathbf{U}}_{L_1}^{[\mathbf{o}^{(1)}]_p}$ is defined as the $N^{(1)}_sL_1 \times N^{(1)}_s$ 0-1 matrix with entries
\begin{equation}\label{eq10}
{\left[ {{\mathbf{U}}_{L_1}^{{[\mathbf{o}^{(1)}]_p}}} \right]_{k,m}} = \left\{ {\begin{array}{*{20}{c}}
  1&{{\text{if mod}}\left( {k - 1,L_1} \right) = {[\mathbf{o}^{(1)}]_p}} \\ 
  0&{{\text{otherwise}}} 
\end{array}} \right.,
\end{equation} 
where the vector ${\mathbf{o}^{(1)}} = \left[ {\begin{array}{cccc} {{o^{(1)}_0}}& o^{(1)}_1 & \cdots &{{o^{(1)}_{P - 1}}} \end{array}} \right]$ contains the time offsets per parallel data frame.
The output of the upsampling operation per parallel stream is then given by the matrix product ${\mathbf{U}}_{L_1}^{[\mathbf{o}^{(1)}]_p}{\mathbf{C}^{(1)}\mathbf{X}_p\mathbf{E}^{(1)}}$. To incorporate all $P$ parallel streams, one can define the block diagonal $PL_1N^{(1)}_s \times PN^{(1)}_s$ matrix ${\mathbf{U}^{(1)}_{\mathrm{ext}}} = \text{diag}({{\mathbf{U}}_{L_1}^{{[\mathbf{o}^{(1)}]_0}}},{{\mathbf{U}}_{L_1}^{{[\mathbf{o}^{(1)}]_1}}},\ldots,{{\mathbf{U}}_{L_1}^{{[\mathbf{o}^{(1)}]_{P - 1}}}})$. The upsampled signal can then be written as
\begin{equation}\label{eq11}
\mathrm{upsampling:\ } {\mathbf{U}^{(1)}_{\mathrm{ext}}}{{\mathbf{C}^{(1)}_{\mathrm{ext}}}}{\mathbf{X}^{(1)}}{{\mathbf{E}^{(1)}}}
\end{equation}

Circular convolution can be represented by the $N^{(1)}_{\mathrm{c}} \times N^{(1)}_{\mathrm{c}}$ matrix of~(\ref{eq12}) (see top of the next page), where it was asumed that (as usual) all the subchannels employ the same filter, $h$. 
\begin{figure*}
\begin{equation}\label{eq12}
{\mathbf{H}^{(1)}} = \left[ {\begin{array}{*{20}{c}}
  {{h_0}}&0& \ldots &0&{{h_{{K_0} - 1}}}&{{h_{{K_0} - 2}}}& \ldots &{{h_1}} \\ 
  {{h_1}}&{{h_0}}&0& \ldots &0&{{h_{{K_0} - 1}}}& \ldots &{{h_2}} \\ 
   \ddots & \ddots & \ddots & \ddots & \ddots & \ddots & \ddots & \ddots  \\ 
  {{h_{{K_0} - 1}}}&{{h_{{K_0} - 2}}}& \ldots &{{h_1}}&{{h_0}}&0& \ldots &0 \\ 
  0&{{h_{{K_0} - 1}}}&{{h_{{K_0} - 2}}}& \ldots &{{h_1}}&{{h_0}}&0& \ldots  \\ 
   \ddots & \ddots & \ddots & \ddots & \ddots & \ddots & \ddots &{} \\ 
   \ddots & \ddots & \ddots & \ddots & \ddots & \ddots & \ddots & \ddots  \\ 
  0& \ldots &0&{{h_{{K_0} - 1}}}&{{h_{{K_0} - 2}}}& \ldots &{{h_1}}&{{h_0}} 
\end{array}} \right]
\end{equation}
\hrulefill
\end{figure*}
Nevertheless, in general, the upsampled signal of~(\ref{eq11}) has not necessarily dimensions compatible with those of $\mathbf{H}^{(1)}$. Four cases can be identified:
\begin{itemize}
\item $N^{(1)}_\mathrm{c} = L_1N^{(1)}_s$: circular convolution is performed with circular filtering period equal to the signal length.
\item $L_1N^{(1)}_s<N^{(1)}_\mathrm{c}<L_1N^{(1)}_s+K_0-L_1+\mathsf{max}(\mathbf{o}^{(1)})$: circular convolution is performed, however the circular filtering period is longer than the signal duration and zero stuffing is necessary. The term $\mathsf{max}(\mathbf{o}^{(1)})$ indicates the element of $\mathbf{o}^{(1)}$ with the maximum value.
\item $N^{(1)}_\mathrm{c} \ge L_1N^{(1)}_s+K_0-L_1+\mathsf{max}(\mathbf{o}^{(1)})$: filtering leads to linear convolution. Zero stuffing is also necessary.
\item $N^{(1)}_\mathrm{c}<L_1N^{(1)}_s$: information loss is unavoidable. This case cannot be met in a functional modulation scheme.
\end{itemize} 
Thus, if $\mathbf{Z}^{(1)} = \left[ {\begin{array}{*{20}{c}}
  {{{\mathbf{I}}_{L_1{N^{(1)}_s}}}} \\ 
  {{{\mathbf{0}}_{\left( {{N^{(1)}_c} - L_1{N^{(1)}_s}} \right) \times L_1{N^{(1)}_s}}}} 
\end{array}} \right]$ is the zero stuffing matrix, then the convolution result is given by
\begin{equation}\label{eq13}
{\text{convolution~result:}} {~\mathbf{H}^{(1)}\mathbf{Z}^{(1)}\mathbf{U}_{L_1}^{{{[{\mathbf{o}^{(1)}}]}_p}}}{\mathbf{C}^{(1)}}{{\mathbf{X}}_p}{\mathbf{E}^{(1)}}
\end{equation}

Up to this point, all the subchannels have been filtered with the prototype filter but no modulation to the proper frequency domain subchannel has been performed. The mapping of~(\ref{eq9}) in a compact matrix relationship is a cumbersome task. To this end, the following ${N^{(1)}_\mathrm{c}} \times M_1$ DFT-like modulation matrix is defined: 
\begin{equation}\label{eq14}
	{\left[ {\mathbf{F}^{(1)}_{b^{(1)}_{\text{conj}}}} \right]_{k,m}} = \exp \left[ {2\pi j\frac{{(k - 1)(m - 1)}}{M}(-1)^{b^{(1)}_{\text{conj}}}} \right]
\end{equation}
where $k=1,2,\ldots,N^{(1)}_\mathrm{c}$, $m=1,2,\ldots,M_1$. The parameter $b^{(1)}_{\text{conj}}$ is introduced to accommodate conjugation or no conjugation for matrix $\mathbf{F}^{(1)}$. If $b^{(1)}_{\text{conj}}=1$, the modulator extracts the conjugate result regarding the modulation of subchannels. This is necessary for the implementation of the Single Carrier (SC) extensions of the multicarrier modulation schemes. For example, SC-FDMA (used, for example, in LTE uplink) employs a DFT operation at the first stage and an IDFT operation at the second stage. The parameter ${b^{(1)}_{{\text{conj}}}}$ will allow the integration of the SC modulation extensions in the framework under development. 

One can then verify that the following expression is equivalent with~(\ref{eq9}):
\begin{equation}\label{eq15}
{{\mathbf{X'''}}_p} = {\mathbf{F}^{(1)}_0} \circ \left( {{\mathbf{H}^{(1)}}\left( {{{\mathbf{F}^{(1)}_0}^*} \circ \left( {{\mathbf{Z}^{(1)}\mathbf{U}_{L_1}^{{{[{\mathbf{o}^{(1)}}]}_p}}}{{{\mathbf{X''}}}_p}} \right)} \right)} \right)
\end{equation}
where $b^{(1)}_{\text{conj}}=0$ is assumed, as in the conventional filtering case. The reader should notice that the associativity property does not hold in~(\ref{eq15}). It must also be noted that, in the typical case where 
$L_1=M_1$, the inner Hadamard product with $\mathbf{F}^{(1)*}_{b^{(1)}_{\text{conj}}}$ is omitted. 

In order to extend the result to the $P$ parallel stream model, the following extended matrices are defined:
\begin{equation}\label{eq16}
\begin{gathered}
  {{\mathbf{Z}}^{(1)}_\mathrm{ext}} = \mathbf{I}_{P}\otimes \mathbf{Z}^{(1)} \hfill \\
  {{\mathbf{F}}^{(1)}_\mathrm{ext}} = \mathbf{1}_{P}\otimes \mathbf{F}^{(1)}_{b_\text{conj}} \hfill \\
  {{\mathbf{H}}^{(1)}_\mathrm{ext}} = \mathbf{I}_{P}\otimes \mathbf{H}^{(1)} \hfill
\end{gathered}
\end{equation}
where $\mathbf{Z}^{(1)}_\mathrm{ext}$ and $\mathbf{H}^{(1)}_\mathrm{ext}$ are block diagonal. Thus:
\begin{equation}\label{eq17}
{\mathbf{X'''}} = {\mathbf{F}^{(1)}_\mathrm{ext}} \circ \left( {{{\mathbf{H}}^{(1)}_\mathrm{ext}}\left( {{{\mathbf{F}^{(1)}_\mathrm{ext}}^*} \circ \left( {{{\mathbf{Z}}^{(1)}_\mathrm{ext}}{{\mathbf{U}}^{(1)}_\mathrm{ext}}{\mathbf{X''}}} \right)} \right)} \right)
\end{equation}

\subsection{Phase shift}
\label{SubSec22}

The phase shift, that has to be compensated, results from the use of causal filters. A Boolean (0-1) parameter ${b^{(1)}_{{\text{cas}}}}$ is introduced to represent this. If ${b^{(1)}_{{\text{cas}}}}=1$, then the phase shift will be compensated, yielding the non-causal filter result. 
The parameter ${b^{(1)}_{{\text{cas}}}}$ is applied directly in (\ref{eq9}) as follows:
\begin{equation}\label{eq18}
\bar x_{p,{{[{\mathbf{e}}]}_m},n}^{(i)} = {e^{ - \frac{{2\pi j{{[{\mathbf{e}}]}_m}\left( {{K_0} - 1} \right){b^{(1)}_{{\text{cas}}}}}}{{2M_1}}}}{x'''}_{p,{{[{\mathbf{e}}]}_m},n}^{(i)}
\end{equation} 
Incorporation in the matrix formulation of the generic modulator can be achieved with the following formula:
\begin{equation}\label{eq19}
{\mathbf{\bar X}} = \left( {{\mathbf{X'''}}} \right){\text{diag}}({\mathbf{c}^{(1)}})
\end{equation}
where $\mathbf{c}^{(1)}$ is the $M_1 \times 1$ vector with 
\[{\left[ {\mathbf{c}^{(1)}} \right]_m} = {e^{ - \frac{{\pi j\left( {{K_0} - 1} \right)(m - 1){b^{(1)}_{\text{cas}}}}}{M_1}}},\;m = 0,1,\ldots,M_1 - 1 \]

\subsection{Downsampling}
\label{SubSec23}

The next step of the filtering process is the signal downsampling by $Q_1$. Similarly to upsampling, a sample offset may be applied during decimation. Therefore a $P \times 1$ vector $\mathbf{a}^{(1)}$ that contains the sample offset per parallel stream is defined in a similar way with $\mathbf{o}^{(1)}$. The decimation result is given by:
\begin{equation}\label{eq20}
{\tilde x}_{p,{{[{\mathbf{e}}]}_m},n}^{(i)} = \bar x_{p,{{[{\mathbf{e}}]}_m},Q_1n + {{\left[ {\mathbf{a}}^{(1)} \right]}_p}}^{(i)}
\end{equation}
Again, to express the decimation for the $p$th stream in a matrix form, the $\left\lfloor {\frac{{{N^{(1)}_c}}}{Q}} \right\rfloor  \times {N^{(1)}_c}$ decimation matrix ${\mathbf{D}}_{Q_1}^{[\mathbf{a}^{(1)}]_p}$ is defined as follows
\begin{equation}\label{eq21}
{\left[ {{\mathbf{D}}_{Q_1}^{{ [\mathbf{a}^{(1)}]_p}}} \right]_{k,m}} = \left\{ {\begin{array}{*{20}{c}}
  1&{{\text{if mod}}\left( {m - 1,Q_1} \right) = {[\mathbf{a}^{(1)}]_p}} \\ 
  0&{{\text{otherwise}}} 
\end{array}} \right.
\end{equation}
For the extension to $P$ parallel streams, a block diagonal downsampling matrix 
\[
\mathbf{D}^{(1)}_\mathrm{ext} = \text{diag}({{\mathbf{D}}_{Q_1}^{{ [\mathbf{a}^{(1)}]_0}}},{{\mathbf{D}}_{Q_1}^{{ [\mathbf{a}^{(1)}]_1}}},\ldots,{{\mathbf{D}}_{Q_1}^{{ [\mathbf{a}^{(1)}]_{P-1}}}} )
\]
is applied and the result is
\begin{equation}
{\mathbf{\tilde X}} = {{\mathbf{D}}^{(1)}_\mathrm{ext}}{\mathbf{\bar X}}
\end{equation}
At this point, the size of the signal matrix ${\mathbf{\tilde X}}_p$ of the $p$th parallel stream and $i$th frame is $\left\lfloor {\frac{{{N^{(1)}_{\mathrm{c}}}}}{Q_1}} \right\rfloor  \times M_1$.

\subsection{Second tier of cyclic extension }
\label{SubSec24}

It is possible that the cyclic extension is not applied directly in the incoming data streams but in the result of the first filtering block. A typical example is the conventional OFDM scheme, where the IDFT block of the OFDM modulator plays the role of the first filtering stage.
Similarly with Section~\ref{SubSec11}, the parameters $N_{\text{cp}}^{(2)}$, $N_{\text{cs}}^{(2)}$, $N_{\text{zp}}^{(2)}$, $N_{\text{zs}}^{(2)}$ are defined and $N_s^{(2)} = N_{\text{cp}}^{(2)}+N_{\text{cs}}^{(2)}+N_{\text{zp}}^{(2)}+N_{\text{zs}}^{(2)}+N'$, where $N'=\left\lfloor {\frac{{{N^{(1)}_c}}}{Q_1}} \right\rfloor$. The extended signal will be given by~(\ref{eq22}) (see top of this page).
\begin{figure*}
\begin{equation}\label{eq22}
y_{p,m,n}^{(i)} = \left\{ \begin{array}{ll}
  0, & n = - N_{\text{zp}}^{(2)} - N_{\text{cp}}^{(2)},\ldots,- N_{\text{cp}}^{(2)} - 1 \\ 
  \tilde{x}_{p,m,\bmod(n,N')}^{(i)}, & n =  - N_{\text{cp}}^{(2)},\ldots,N' + N_{\text{cs}}^{(2)}-1 \\ 
  0, & n = N' + N_{\text{cs}}^{(2)},\ldots,N' + N_{\text{cs}}^{(2)} + N_{\text{zs}}^{(2)} - 1
\end{array} \right.
\end{equation}
\hrulefill
\end{figure*}
Moreover, in a manner analogous with~(\ref{eq4}), one can define the matrix $\mathbf{C}^{(2)}$ 
\begin{equation}\label{eq23}
{{\mathbf{C}}^{(2)}} = \left[ {\begin{array}{*{20}{c}}
  {{{\mathbf{0}}_{N_{\text{zp}}^{(2)} \times N'}}} \\ 
  {\begin{array}{*{20}{c}}
  {{{\mathbf{0}}_{N_{\text{cp}}^{(2)} \times \left( {N' - N_{\text{cp}}^{(2)}} \right)}}}&{{{\mathbf{I}}_{N_{\text{cp}}^{(2)}}}} 
\end{array}} \\ 
  {{{\mathbf{I}}_{N'}}} \\ 
  {\begin{array}{*{20}{c}}
  {{{\mathbf{I}}_{N_{\text{cs}}^{(2)}}}}&{{{\mathbf{0}}_{N_{\text{cs}}^{(2)} \times \left( {N' - N_{\text{cs}}^{(2)}} \right)}}} 
\end{array}} \\ 
  {{{\mathbf{0}}_{N_{\text{zs}}^{(2)} \times N'}}} 
\end{array}} \right]
\end{equation}
and its corresponding block diagonal extension ${\mathbf{C}}_{\mathrm{ext}}^{(2)}$. The signals can then be written as
\begin{equation}\label{eq24}
\begin{gathered}
  {{\mathbf{Y}}_p} = {{\mathbf{C}}^{(2)}}{{{\mathbf{\tilde X}}}_p} \hfill \\
  {\mathbf{Y}} = {\mathbf{C}}_{\mathrm{ext}}^{(2)}{\mathbf{\tilde X}} \hfill \\ 
\end{gathered}
\end{equation}

\section{Windowing}

In the windowing block, a time-domain window is applied in the signal. For example, in Windowed OFDM (used, e.g., in ITU Ghn~\cite{ghn}), a window is applied in the OFDM waveform to suppress its sidelobes. The application of the time-domain window can be expressed as follows
\begin{equation}\label{eq25}
\bar y_{p,{{[{\mathbf{e}}]}_m},n}^{(i)} = {w_n}y_{p,{{[{\mathbf{e}}]}_m},n}^{(i)},
\end{equation}
where $w_n$ is the time-domain window, of $N_s^{(2)}$ samples. If the defined window length is smaller, then the window should be properly zero-padded. If no window is considered in the modulation scheme, then $w_n=1$ for all $n$.

To express this in a matrix form, a diagonal matrix with the window samples on its diagonal is defined, that is, 
\[{\mathbf{W}} = \left[ {\begin{array}{*{20}{c}}
  {{w_0}}&0& \ldots &0 \\ 
  0&{{w_1}}& \ddots & \vdots  \\ 
   \vdots & \ddots & \ddots &0 \\ 
  0& \ldots &0&{{w_{N_s^{(2)} - 1}}} 
\end{array}} \right]\] 
The windowed signal is then given by
\begin{equation}\label{eq26}
\begin{gathered}
  {{{\mathbf{\bar Y}}}_p} = {\mathbf{W}}{{\mathbf{Y}}_p} \hfill \\
  {\mathbf{\bar Y}} = {{\mathbf{W}}_{\mathrm{ext}}}{\mathbf{Y}} \hfill \\ 
\end{gathered} 
\end{equation}
where ${{\mathbf{W}}_{\mathrm{ext}}} = \mathbf{I}_{P}\otimes {\mathbf{W}}$.

\section{Second Filtering Stage and Third Tier of Cyclic Extension}

The need for a second generalized filtering stage is imposed by the will to include all the popular multicarrier modulations, as well as popular SC extensions. For example, since UFMC~\cite{6824990} is practically a shaped OFDM waveform, a second filtering stage is required to accommodate it in the model. Accordingly, SC modulations, such as SC-FDMA and SC-FBMC, can also be modeled with the introduction of a second filtering stage. In the following analysis, it is assumed that the second filtering stage consists of $M_2$ subchannels.

\subsection{Multiplexer}

The multiplexer is the block that combines and routes the output of the $M$ subchannels of the first filtering stage to the second one. The multiplexer can play various roles in the modulation chain:
\begin{itemize}
\item {it directs the output of subchannels of the first filtering stage to subchannels of the second filtering stage}
\item {it combines the output of multiple subchannels of the first filtering stage}
\item {it transposes the output of the first filtering stage, i.e. it serializes the output of the subchannels of the first filter}
\item {it may substitute the role of the decimator with the incorporation of downsampling into the routing/combining procedures} 
\end{itemize}  

Typically, in order to model the operation of a multiplexer, an $M_1 \times M_2$ 0-1 matrix ${{\mathbf{E}}^{(2)}}$ can be defined as follows. If ${\left[ {{{\mathbf{E}}^{(2)}}} \right]_{{m_1},{m_2}}} = 1$, then the output of the $m_1$th subchannel of the first bank is fed to the $m_2$th subchannel of the second bank. It is also allowed to feed multiple subchannels of the first stage to the same subchannel of the second stage (signal multiplexing). This means that the input for each subchannel of the second filtering stage is given by:
\begin{equation}\label{eq27}
r_{p,n,m_2}^{(i)} = \sum\limits_{{m_1} = 0}^{{M_1} - 1} {{{\left[ {\mathbf{E}^{(2)}} \right]}_{{m_1},{m_2}}} {\bar y}_{p,{m_1},n}^{(i)}}
\end{equation}
Multiplexing can be also expressed in matrix form with a simple post-multiplication, namely,
\begin{equation}\label{eq28}
\begin{gathered}
  {{\mathbf{R}}_p} = {{{\mathbf{\bar Y}}}_p}{{\mathbf{E}}^{(2)}}, \hfill \\
  {\mathbf{R}} = {\mathbf{\bar Y}}{{\mathbf{E}}^{(2)}} \hfill \\ 
\end{gathered} 
\end{equation}
The sole exception not covered by (\ref{eq28}) is related to the third role of the multiplexer and it appears in the SC extensions of the multicarrier modulation schemes. In SC extensions, the output of the first stage should be multiplexed and transposed before fed to the second filter. In order to express this procedure in a single relationship, the boolean parameter $b_\text{tran}$ is introduced. If $b_\text{tran}=1$, then the output of the first filtering stage will be transposed. In this case, due to the transpose operation, the matrix ${{\mathbf{E}}^{(2)}}$ performs practically pre-multiplication of the output. Therefore, an additional auxiliary 0-1 multiplexing matrix ${{\mathbf{E}}^{(3)}}$ is considered for post-multiplication. 
\begin{equation}\label{eq29}
{{\bf{R}}_p} = \left( {1 - {b_{{\rm{tran}}}}} \right){{{\bf{\bar Y}}}_p}{{\bf{E}}^{(2)}} + {b_{{\rm{tran}}}}{\left( {{{{\bf{\bar Y}}}_p}{{\bf{E}}^{(2)}}} \right)^T}{{\bf{E}}^{(3)}}
\end{equation}

\subsection{Second generalized filtering stage}

The second filtering stage includes all the functional blocks that were described in Section~\ref{Sec2} and hence the analysis will not be repeated here. Given that the input of the second filtering stage is provided by~(\ref{eq27}) and~(\ref{eq28}), its output can be easily written using the proper parameters ($L_{2},~Q_{2}, \mathbf{H}^{(2)}$, etc.). The result of the second generalized filtering stage that combines the analysis of Section~\ref{Sec2} with~(\ref{eq23}), (\ref{eq26}), and~(\ref{eq28}) is given in~(\ref{eqYp2}) (at the top of the next page).
\begin{figure*}
\begin{equation}
\label{eqYp2}
{\mathbf{Y}}_p^{(2)} = {\mathbf{D}}_{{Q_2}}^{{{[{{\mathbf{a}}^{(2)}}]}_p}}\left( {{{\mathbf{F}}^{(2)}_{b^{(2)}_\text{conj}}} \circ \left( {{\mathbf{H}}{}^{(2)}\left( {{{\mathbf{F}}^{(2)*}_{b^{(2)}_\text{conj}}} \circ \left( {{{\mathbf{Z}}^{(2)}}{\mathbf{U}}_{{L_2}}^{{{[{{\mathbf{o}}^{(2)}}]}_p}}{\mathbf{W}}{{\mathbf{C}}^{(2)}}{{{\mathbf{R}}}_p}} \right)} \right)} \right)} \right)\text{diag}\left( {{{\mathbf{c}}^{(2)}}} \right)
\end{equation}
\hrulefill
\end{figure*}

\subsection{Third tier of cyclic extension}
If desired, a third stage of cyclic extension can be added. The procedure and the definition of the analogous matrix $\mathbf{C}^{(3)}$ are identical with those described in Sections~\ref{SubSec11} and~\ref{SubSec24}.

\section{Modulator Output}

The second filtering stage yields a set of $PM_2$ signals. Thus, another multiplexing block is required to combine those extracted signals. The outputs of all $M_2$ frequency domain subchannels should be added. Therefore:
\begin{equation}
  {{\mathbf{S}}_p} = {\mathbf{Y}}_p^{(2)}\mathbf{1}_{M_2}
\end{equation}
where ${\mathbf{Y}}_p^{(2)}$ denotes the output of the second filtering stage.

Regarding the $P$ parallel streams, another multiplexing stage should be introduced. The multiplexer will be given by a $KN'' \times PN''$ matrix, where $N''$ is the signal length per parallel stream at the output of the second filtering stage and $K$ is the number of outputs of the modulator. For example, in multiple-input multiple-output (MIMO) transmission with spatial multiplexing of $P$ layers, $K=P$ and the multiplexer will be the identity. For single-antenna FBMC-OQAM, $P=2$ and $K=1$. As a consequence, the multiplexer output is given by 
\begin{equation}
  \mathbf{S} = \left[ {\begin{array}{*{20}{c}}
  {{{\mathbf{I}}_{N''}}}&{{{\mathbf{I}}_{N''}}} 
\end{array}} \right]\left[ {\begin{array}{*{20}{c}}
  {{{\mathbf{S}}_0}} \\ 
  {{{\mathbf{S}}_1}} 
\end{array}} \right] ~~(\text{Final Output}) 
\end{equation}
that is, the multiplexer matrix is in this case
\[
{{\mathbf{E}}^{(4)}} =  \left[ {\begin{array}{*{20}{c}}
  {{{\mathbf{I}}_{N''}}}&{{{\mathbf{I}}_{N''}}} 
\end{array}} \right]
\]

\section{Conclusions}

A unification of the single- and multi-carrier modulation formats was presented, in a structured matrix framework. With appropriate parameter settings, the generic modulator developed in this paper can provide a given modulation scheme as a special case. For example, the parametrizations needed to generate the CP-OFDM, FBMC-OQAM, and SC-FDMA schemes are outlined in Table~\ref{1234}. 
\begin{table*}[h]
\caption{Parameterization of multicarrier modulations}
\label{1234}
\begin{tabular}{|C{2.5cm}|C{5.2cm}|C{4.5cm}|C{5.2cm}|}
\hline
 & OFDM (fully loaded 128 subcarriers, 0.25 cyclic prefix)     & FBMC-OQAM  (fully loaded 32 subchannels, 200 time symbols per frame, no cyclic prefix) & SC-FDMA (128 subcarriers, 0.25 cyclic prefix, using quarter of available bandwidth)    \\
 \hline
$N$    				& 1    & 200    & 1    \\
$M' $  				& 128  & 32     & 32   \\
$M_1$  				& 128  & 32     & 32   \\
$L_1$  				& 128  & 32     & 32   \\
$Q_1$  				& 1    & 1      & 1    \\
$P$    				& 1    & 2      & 1  \\
$\mathbf{e}$	    & $\left[ {\begin{array}{*{20}{c}}0&1& \cdots &{{M_1} - 1}\end{array}} \right]$	   & 	$\left[ {\begin{array}{*{20}{c}}0&1& \cdots &{{M_1} - 1}\end{array}} \right]$	& 	$\left[ {\begin{array}{*{20}{c}}0&1& \cdots &{{M_1} - 1}\end{array}} \right]$	\\

$N^{(1)}_{cp}$		& 0    & 0      & 0  \\
$N^{(1)}_{cs}$    	& 0    & 0      & 0  \\
$N^{(1)}_{zp}$   	& 0    & 0      & 0  \\
$N^{(1)}_{zs}$    	& 0    & 0      & 0  \\
$h$					& $\scriptstyle{h_n^m = \left\{ {\begin{array}{*{20}{c}}
{1,n = 0,\ldots,{M_1} - 1}\\
{{\rm{else~~0}}}
\end{array},{m = 0,\ldots,{M_1} - 1}} \right.}$	   & A proper well-localized filter & 	$\scriptstyle{h_n^m = \left\{ {\begin{array}{*{20}{c}}
{1,n = 0,\ldots,{M_1} - 1}\\
{{\rm{else~~0}}}
\end{array},m = 0,\ldots,{M_1} - 1} \right.}$ \\
$\mathbf{o}^{(1)} $	 &	0			& ${\left[ {\begin{array}{*{20}{c}}
0&{{{{M_1}} \mathord{\left/
 {\vphantom {{{M_1}} 2}} \right.
 \kern-\nulldelimiterspace} 2}}
\end{array}} \right]^T} $ & 0\\
$\mathbf{a}^{(1)} $		& 0 & 0 & 0 \\
$b^{(1)}_{\text{conj}}$ 			&	0			&  0		& 	1\\
$b^{(1)}_{\text{cas}}$ 			&	0			&  1		& 	0\\ 

$N^{(2)}_{cp}$		& $0.25M_1$     & 0			   		& 0  \\
$N^{(2)}_{cs}$    	& 0    			& 0      			& 0  \\
$N^{(2)}_{zp}$   	& 0    			& 0      			& 0  \\
$N^{(2)}_{zs}$    	& 0    			& 0      			& 0  \\
$b_\text{tran}$		& 0				& 0					& 1  \\
$\mathbf{E}^{(2)}$  &	$\mathbf{1}_{M_1}^{T}$			&	$\mathbf{1}_{M_1}^{T}$				&  $\mathbf{1}_{M_1}^{T}$ \\
$\mathbf{E}^{(3)}$  & $[~]$	     	& $[~]$	&	$\left[ {\begin{array}{*{20}{c}}
{{{\bf{0}}_{{M_1} \times \left( {{M_2} - {M_1}} \right)}}}&{{{\bf{I}}_{{M_1}}}}
\end{array}} \right]$  \\

$M_2$  				& 1		  		& 1  	 	   		& 128   \\
$L_2$  				& 1 	  		& 1 	       		& 128   \\
$Q_2$  				& 1   	  		& 1  	       		& 1     \\
$g$					& $\delta(n)$	& $\delta(n)$		   &  $\scriptstyle {g_n^m = \left\{ {\begin{array}{*{20}{c}}
{1,n = 0,\ldots,{M_2} - 1}\\
{{\rm{else~~0}}}
\end{array},m = 0,\ldots,{M_2} - 1} \right.}$ 	  \\
$\mathbf{o}^{(2)} $	 &	0			& 0 & 0\\
$\mathbf{a}^{(2)} $		& 0 & 0 & 0 \\
$b^{(2)}_{\text{conj}}$ 			&	0			&  0		& 	0\\
$b^{(2)}_{\text{cas}}$ 			&	0			&  0		& 	0\\ 

$N^{(3)}_{cp}$		& 0		        & 0			   		& $0.25M_2$  \\
$N^{(3)}_{cs}$    	& 0    			& 0      			& 0  \\
$N^{(3)}_{zp}$   	& 0    			& 0      			& 0  \\
$N^{(3)}_{zs}$    	& 0    			& 0      			& 0  \\

$\mathbf{E}^{(4)}$ 	&$\mathbf{I}$	& 	$\left[ {\begin{array}{*{20}{c}}
{{{\bf{I}}_{N''}}}&{{{\bf{I}}_{N''}}}
\end{array}} \right]$	& 	$\mathbf{I}$  \\

$\mathbf{W}$	&  $\mathbf{I}_{M_1(1+0.25)}$		& 	$\mathbf{I}_{L_1N+K_0-L_1/2}$		& $\mathbf{I}_{M_1}$\\

Input			& Arranged in $M' \times 1$ complex symbol blocks		& Input is arranged in $2 \times M \times N$ blocks each carrying the real and imaginary part of the initial QAM symbols as in (\ref{eq30}).

		& Arranged in $M' \times 1$ complex symbol blocks\\
\hline
\end{tabular}
\begin{equation}\label{eq30}
x_{p,m,n}^{(i)} = \left\{ \begin{array}{ll}
\Re(a_{m,n}^{(i)}), & p = 0 \mbox{\ and\ } m \mbox{\ is even}\\
j\Im(a_{m,n}^{(i)}), & p = 1 \mbox{\ and\ } m \mbox{\ is even}\\
j\Im(a_{m,n}^{(i)}), & p = 0 \mbox{\ and\ } m \mbox{\ is odd}\\
\Re(a_{m,n}^{(i)}), & p = 1 \mbox{\ and\ } m \mbox{\ is odd}
\end{array}\right.
\end{equation}
\hrulefill
\end{table*}

Future work in this context can include: development of a unified model for the demodulator; study of common receiver blocks (e.g., channel estimation / equalization) in this generic framework; facilitation of implementation and comparison of existing multi- and single-carrier schemes, and synthesis of new (or hybrid) ones.

\bibliographystyle{IEEEtran}
\bibliography{IEEEabrv,FBMCunification}

\end{document}